\magnification\magstep1
\font\bf=cmcsc10
\font\smc=cmcsc10

\centerline{\bf Locality and Bell's inequality}

\bigskip\bigskip

\centerline{\bf Luigi Accardi, Massimo Regoli}

\bigskip
\centerline{Centro Vito Volterra }

\centerline{  Universit\`a di Roma ``Tor Vergata'', Roma, Italy }

\centerline{  email: accardi@volterra.mat.uniroma2.it,
              WEB page: http://volterra.mat.uniroma2.it}

\bigskip\bigskip

\bigskip

\centerline{\bf WARNING}

This is a revised version with respect to the one submitted on Mon, 3
Jul 2000. By mistake an earlier version and not the final one was
submitted.

\centerline{\bf Abstract}
We prove that the locality condition is irrelevant to Bell in equality.
We check that the real origin of the Bell's inequality is the
assumption of applicability of classical (Kolmogorovian) probability
theory to quantum mechanics. We describe the chameleon effect which
allows to construct an experiment realizing a local, realistic, classical,
deterministic and macroscopic violation of the Bell inequalities.

Index:
-- Inequalities among numbers

-- The Bell inequality

-- Implications of the Bell's inequalities for the singlet correlations

-- Bell on the meaning of Bell's inequality

-- Critique of Bell's ``vital assumption''

-- The role of the counterfactual argument in Bell's proof

-- Proofs of Bell's inequality based on counting arguments

-- The quantum probabilistic analysis

-- The realism of ballot boxes and the corresponding statistics

-- The realism of chameleons and the corresponding statistics

-- Bell's inequalities and the chamaleon effect

-- Physical implausibility of Bell's argument

-- The role of the single probability space in CHSH's proof

-- The role of the counterfactual argument in CHSH's proof

-- Physical difference between the CHSH's and the original Bell's inequalities

-- Bibliography
\bigskip

\beginsection{(1) Inequalities among numbers }

\bigskip
\noindent{\bf Lemma (1)} For any two numbers $a,c\in[-1,1]$ the following equivalent
inequalities hold:
$$|a \pm c|\leq1 \pm ac\eqno(1)$$
Moreover equality in (1) holds if and only if either $a=\pm 1$ or $c=\pm 1$.
\bigskip
\noindent{\bf Proof}. The equivalence of the two inequalities (1) follows
from the fact that one is obtained from the other by changing the sign
of $c$ and $c$ is arbitrary in $[-1,1]$.

Since for any $a,c\in[-1,1]$, $1 \pm ac\geq0$, (1) is equivalent to
$$|a \pm c|^2=a^2+c^2 \pm 2ac\leq(1 \pm ac)^2=1+a^2c^2 \pm 2ac$$
and this is equivalent to
$$a^2(1-c^2)+c^2\leq1$$
which is identically satisfied because $1-c^2\geq0$ and therefore
$$a^2(1-c^2)+c^2\leq1-c^2+c^2=1\eqno(2)$$
Notice that in (2) equality holds if and only if $a^2=1$ i.e. $a=\pm 1$.
Since, exchanging $a$ and $c$ in (1) the inequality remains unchanged,
the thesis follows.
\bigskip
\noindent{\bf Corollary (2)}
For any three numbers $a,b,c\in[-1,1]$ the following
equivalent inequalities hold:
$$|ab \pm cb|\leq1 \pm ac\eqno(3)$$
and equality holds if and only if $b=\pm 1$ and either $a=\pm 1$ or $c=\pm 1$.
\bigskip
\noindent{\bf Proof.} For $b\in[-1,1]$,
$$|ab \pm cb|=|b|\cdot|a \pm c|\leq|a \pm c|\eqno(4)$$
so the thesis follows from Lemma (1). In (4) the first equality holds if
and only if $b=\pm 1$, so also the second statement follows from Lemma (1).
\bigskip
\noindent{\bf Lemma (3)}. For any numbers $a$, $a'$, $b$, $b'$, $c\in[-1,1]$,
one has
$$|ab-bc|+|ab'+b'c|\leq2\eqno(5)$$
$$ab+ab'+a'b'-a'b\leq2\eqno(6)$$
In (5) equality holds if and only if $b,b',a, c=\pm 1$.
\bigskip
\noindent{\bf Proof}. Because of (3)
$$|ab - cb|\leq1 - ac\eqno(7)$$
$$|ab' - cb'|\leq1 + ac\eqno(8)$$
adding (7) and (8) one finds (5). The left hand side of (6) is $\leq$ than
$$|ab-ba'|+|ab'+b'a'|\eqno(9)$$
and replacing $a'$ $by$ $c$, (8) becomes the left hand side of (5).
If $b,b'=\pm 1$ and $a=\pm 1$ equality holds in (7) and (8) hence in (5).
Conversely, suppose that equality holds in (5) and suppose that either
$|b|<1$ or $|b'|<1$. Then we arrive to the contradiction
$$2=|b|\cdot |a-a'|+|b'|\cdot |a+a'| < |a-a'|+|a+a'| \leq
(1-aa')+(1+aa')=2  \eqno(10)$$
So, if equality holds in (5), we must have $|b|=|b'|=1$. In this case
(5) becomes
$$|a-a'|+|a+a'| =2\eqno(11)$$
and, if either $|a|<1$ or $|a'|<1$, then from Lemma (1) we know that
$$|a-a'|+|a+a'| < (1-aa')+(1+aa')=2  $$
so we must also have $a, a'=\pm 1$.
\bigskip
\noindent{\bf Corollary (4)}. If $a,a',b, b',c\in \{-1,1\}$, then
the inequalities (3) and (5) are equivalent and equality holds in all of
them. However the inequality in (6) may be strict.
\bigskip
\noindent{\bf Proof}.
From Lemma (1) we know that the inequalities (1) and (2) are equivalent.
From Lemma (3) we know that (1) implies (5). Choosing $b'=a$ in (5),
since $a=\pm 1$, Lemma (2) implies that (5) becomes
$$|ab -cb|\leq1 - ac$$
which is equivalent to (1).
(6) is equivalent to
$$a(b+b') + a'(b'-b)\leq2\eqno(12)$$
In our assumptions either $(b+b')$ or $(b'-b)$ is zero, so (12) is
either equivalent to
$$a(b+b') \leq2$$
or to
$$a'(b'-b) \leq2$$
and in both cases we can choose $a,b,b'$ or $a',b,b'$ so that the
product is negative and the inequality is strict.
\bigskip

\beginsection{(2) The Bell inequality }

\bigskip
\noindent{\bf Corollary (1) } (Bell inequality)
Let $A,B,C,D$ be random variables  defined on the same probability space
$(\Omega, {\cal F}, P)$ and with values in the interval $[-1,1]$. Then
the following inequalities hold:
$$ E(|AB-BC|)\leq 1-E(AC) \eqno(1)$$
$$ E(|AB+BC|)\leq 1+E(AC) \eqno(2)$$
$$E(|AB-BC|)+E(|AD+DC|)\leq2    \eqno(3)$$
where $E$ denotes the expectation value in the probability space of the
four variables. Moreover (1) is equivalent to (2) and,
if either $A$ or $C$ has values $\pm 1$, then the three
inequalities are equivalent.
\bigskip
\noindent{\bf Proof.} Lemma (1.1) implies the following inequalities
(interpreted pointwise on $\Omega$):
$$|AB-BC|\leq 1-AC$$
$$|AB+BC|\leq 1+AC$$
$$|AB-BC|+|AD+DC|\leq2$$
from which (1), (2), (3) follow by taking expectation and
using the fact that $|E(X)|\leq E(|X|)$. The equivalence
is established by the same arguments as in Lemma (1.1).
\bigskip
\noindent{\bf Remark (2)}. Bell's original proof, as well as the almost totality of
the available proofs of Bell's inequality, deal only with the case of
random variables assuming only the values $+1$ and $-1$. The present
generalization is not without interest because it dispenses from the
assumption that the classical random variables, used to describe quantum
observables, have the same set of values of the latter ones: a hidden
variable theory is required to reproduce the results of quantum theory
only when the hidden parameters are averaged over.
\bigskip
\noindent{\bf Theorem (3)}. Let $S_a^{(1)} , S_c^{(1)} , S_b^{(2)} , S_d^{(2)}$ be
random variables defined on a probability space
$(\Omega, {\cal F}, P)$ and with values in the interval $[-1,+1]$. Then
the following inequalities holds:
$$\left| E(S_a^{(1)}S_b^{(2)}) - E(S_c^{(1)}S_b^{(2)})\right| \leq 
1- E(S_a^{(1)}S_c^{(1)}) \eqno(4)$$
$$\left| E(S_a^{(1)}S_b^{(2)}) + E(S_c^{(1)}S_b^{(2)})\right| \leq 
1 + E(S_a^{(1)}S_c^{(1 )}) \eqno(5)$$
$$\left| E(S_a^{(1)}S_b^{(2)}) - E(S_c^{(1)}S_b^{(2)})\right| +
  \left| E(S_a^{(1)}S_d^{(2)}) + E(S_c^{(1)}S_d^{(2)})\right| \leq 2\eqno(6)$$
{\bf Proof.} This is a rephrasing of Corollary (2).
\bigskip

\beginsection{(3) Implications of the Bell's inequalities for the singlet
correlations}

\bigskip
To apply Bell's inequalities to the singlet correlations,
considered in the EPR paradox, it is enough to observe that they imply the
following
\bigskip
\noindent{\bf Lemma (1)} In the ordinary three-dimensional euclidean space
there exist sets of three, unit length, vectors $a$, $b$, $c$, such that it
is not possible to find a probability space $(\Omega, {\cal F}, P)$ and
six random variables $S_x^{(j)}$ ($x=a,b,c$, $j=1,2$) defined on
$(\Omega, {\cal F}, P)$ and with values in the interval $[-1,+1]$,
whose correlations are given by:
$$E(S_x^{(1)}\cdot S_y^{(2)})=-x\cdot y \qquad ; \qquad x,y=a,b,c\eqno(1)$$
where, if $x=(x_1,x_2,x_3)$, $y=(y_1,y_2,y_3)$ are two three-dimensional
vectors, $x\cdot y$ denotes their euclidean scalar product, i.e. the sum
$x_1y_1+x_2y_2+x_3y_3$.
\bigskip
\noindent{\bf Remark.} In the usual EPR--type experiments,
the random variables $S_a^{(j)},S_b^{(j)}, S_c^{(j)}$ represent the spin 
(or polarization) of particle $j$ of a singlet pair along the three directions 
$a,b,c$ in space. The expression in the right-hand side of (1) is the singlet
correlation of two spin or polarization observables, theoretically predicted
by quantum theory and experimentally confirmed by the Aspect-type experiments.
\bigskip
\noindent{\bf Proof.} Suppose that, for any choice of the unit vectors $x=a,b,c$
there exist random variables $S_x^{(j)}$ as in the statement of the Lemma.
Then, using Bell's inequality in the form (2.5) with
$A=S_a^{(1)}$, $B=S_b^{(2)}$, $C=S_c^{(1)}$),
we obtain
$$\left| E(S_a^{(1)}S_b^{(2)})+E(S_b^{(2)}S_c^{(1)})\right| \leq 
1+E(S_a^{(1)}S_c^{(1)})\eqno(2)$$
Now notice that, if $x=y$ is chosen in (1), we obtain
$$E(S_x^{(1)}\cdot S_x^{(2)})=-x\cdot x=-\left\| x\right\| ^2=-1\quad;\qquad
x=a,b,c$$
and, since $\left| S_x^{(1)}S_x^{(2)}\right| =1$ this is possible if and only if $
S_x^{(1)}=-S_x^{(2)}$ $\left( x=a,b,c\right) $ $P$--almost everywhere.
Using this (2) becomes equivalent to:
$$\left| E(S_a^{(1)}S_b^{(2)})+E(S_b^{(2)}S_c^{(1)})\right| \leq 1-E(S_a^{(1)}S_c^{(2)})$$
or, again using (1), to:
$$\left| a\cdot b+b\cdot c\right| \leq 1+a\cdot c  \eqno(3)$$
If the three vectors $a$, $b$, $c$ are chosen to be in the same plane
and such that $a$ is perpendicular to $c$ and $b$ lies between $a$
and $b$, forming an angle $\theta $ with $a$, then the inequality (3) becomes:
$$cos \theta +\sin \theta \leq 1 \qquad ; \quad 0<\theta <\pi /2\eqno(4)$$
But the maximum of the function of $\theta \longmapsto \sin \theta +\cos
\theta $ in the interval $\left[ 0,\pi /2\right] $ is $\sqrt{2} $ (obtained
for $\theta =\pi /4$). Therefore, for $\theta $ close to $\pi /4$, the
left-hand side of (4) will be close to $\sqrt{2} $ which is more that
$1$. In conclusion, for such a choice of the unit vectors $a$, $b$, $c$,
random variables $S_a^{(1)},S_b^{(2)},S_c^{(1)}, S_c^{(2)}$ as in the statement of the Lemma
cannot exist.
\bigskip
\noindent{\bf Definition (2)} A local realistic  model for the EPR (singlet)
correlations is defined by:
\item{(1)} a probability space $(\Omega, {\cal F}, P)$
\item{(2)} for every unit vector $x$, in the three-dimensional euclidean
space, two random variables $S_x^{(1)}, S_x^{(2)}$ defined on $\Omega$ and with
values in the interval $[-1,+1]$ whose correlations, for any $x,y$, are
given by equation (1).
\bigskip
\noindent{\bf Corollary (3)} If $a,b,c$ are chosen so to violate (4) then
a local realistic model for the EPR correlations,
in the sense of Definition (2), does not exist.
\bigskip
\noindent{\bf Proof.}
Its existence would contradict Lemma (1).
\bigskip
\noindent{\bf Remark.} In the literature one usually distinguishes two types of local
realistic models -- deterministic and stochastic ones.
Both are included in Definition (2): the deterministic models are
defined by random variables $S_x^{(j)}$ with values in the set$\{-1,+1\}$;
while, in the stochastic models, the random variables take values in the
interval $[-1,+1]$. The original paper [Be64] was devoted to the
deterministic case. Starting from [Be71] several papers have been
introduced to justify the stochastic models.
We prefer to distinguish the definition of the models from their
justification.
\bigskip

\beginsection{(4) Bell on the meaning of Bell's inequality}

\bigskip
In the last section of [Be66] (submitted before [Be64], but published
after) Bell briefly describes Bohm hidden variable interpretation of
quantum theory underlining its non local character. He then raises the
question: {\it ... that there is no proof that any hidden variable
account of quantum mechanics must have this extraordinary character ...}
and, in a footnote added during the proof corrections, he claims that:
{\it ... Since the completion of this paper such a proof has been found
[Be64].}

In the short Introduction to [Be64], Bell reaffirms the same ideas, namely
that the result proven by him in this paper shows that:
{\it ... any such [hidden variable] theory which reproduces exactly the
quantum mechanical predictions} must have {\it ... a grossly nonlocal
structure}.

The proof goes along the following scheme: Bell proves an inequality
in which, according to what he says (cf. statement after formula (1) in
[Be64]):

 {\it ...  The vital assumption [2] is that the result $B$ for particle $2$
 does not depend on the setting $a$, of the magnet for particle $1$,
 nor $A$ on $b$}.

The paper [2], mentioned in the above statement, is nothing but the Einstein,
Podolsky, Rosen paper [EPR35] and the locality issue is further
emphasized by the fact that he reports the famous Einstein's statement
[Ein49]: {\it ... But on one supposition we should, in my opinion, absolutely
hold fast: the real factual situation of the system $S_2$ is independent
of what is done with the system $S_1$, which is spatially separated from the
former}.

Stated otherwise: according to Bell, Bell's inequality is a consequence
of the locality assumption.

It follows that a theory which violates the above mentioned inequality
also violates {\it ...  the vital assumption} needed, according to Bell,
for its deduction, i.e. locality.

Since the experiments prove the violation of this inequality, Bell
concludes that quantum theory does not admit a local
completion; in particular quantum mechanics is a nonlocal theory.
To use again Bell's words:

\noindent{\it the statistical predictions of quantum mechanics are
incompatible with separable predetermination} ([Be64], p.199). Moreover
this incompatibility has to be understood in the sense that:
{\it in a theory in which parameters are added to quantum mechanics to
determine the results of individual measurements, without changing the
statistical predictions, there must be a mechanism whereby the setting
of one measuring device can influence the reading of another instrument,
howevere remote. Moreover, the signal involved must propagate
instantaneously,...}
\bigskip

\beginsection{(5) Critique of Bell's ``vital assumption''}

\bigskip
An assumption should be considered ``vital'' for a theorem if, without
it, the theorem cannot be proved.

To favor Bell, let us require much less. Namely let us agree to consider
his assumption {\it vital} if the theorem cannot be proved by taking as its
hypothesis the negation of this assumption.

If even this minimal requirement is not satisfied, then we must conclude
that the given assumption has nothing to do with the theorem.

Notice that Bell expresses his locality condition by the requirement 
that {\it the result $B$ for particle $2$} should not depend on 
{\it the setting $a$, of the magnet for particle $1$} (cf. citation in 
the preceeding section). Let us denote ${\cal M}_1$ (${\cal M}_2$)
the space of all possible measurement settings on system 1 (2).
\bigskip
{\bf Theorem (1)}
For each unit vector $x$ in the three dimensional euclidean
space ($x\in {\bf R}^3$, $\mid x \mid =1$) let be given two random variables
$S^{(1)}_x$, $S^{(2)}_x$ (spin of particle 1 (2) in direction $x$), defined on a
space $\Omega$ with a probability $P$ and with values in the $2$--point
set $\{+1,-1\}$. Fix $3$ of these unit vectors $a,b,c$ and suppose that
the corresponding random variables satisfy the following non locality
condition {\bf [violating Bell's {\it vital assumption}]}:
suppose that the probability space $\Omega$ has the following structure:
$$\Omega = \Lambda \times {\cal M}_1\times {\cal M}_2\eqno(1)$$
so that, for some function 
$ F^{(1)}_a,  F^{(2)}_a :\Lambda \times {\cal M}_1\times {\cal M}_2\to [-1,1]$,
$$ S^{(1)}_a(\omega) =  F^{(1)}_a (\lambda,m_1,m_2)
\quad (S^{(1)}_a \ depends \ on \ m_2)\eqno(2)$$
$$ S^{(2)}_a(\omega)  =  F^{(2)}_a (\lambda,m_1,m_2)
\quad (S^{(2)}_a\ depends \ on \ m_1  )\eqno(3)$$
with $m_1\in {\cal M}_1, m_2\in {\cal M}_2$
and similarly for $b$ and $c$. [nothing changes in the proof if we add
further dependences, for example $F^{(2)}_a$ may depend on all the
$S^{(1)}_x(\omega)$ and $F^{(1)}_a$ on all the $ S^{(2)}_x(\omega)$].

Then the random variables $S^{(1)}_a, S^{(2)}_b, S^{(1)}_c$
satisfy the inequality
$$\mid \langle S^{(1)}_a S^{(2)}_b\rangle  - 
\langle  S^{(2)}_b S^{(1)}_c\rangle  \mid \leq
1 - \langle S^{(1)}_a S^{(1)}_c\rangle  \eqno(4)$$
If moreover the singlet condition 
$$\langle S^{(1)}_x \cdot S^{(2)}_x\rangle  = -1\qquad ;\qquad x= a,b,c
\eqno(5)$$
is also satisfied, then Bell's inequality holds in the form
$$\mid\langle S^{(1)}_aS^{(2)}_b\rangle-\langle S^{(2)}_b
S^{(1)}_c\rangle\mid
\leq 1 + \langle S^{(1)}_a S^{(2)}_c\rangle  \eqno(6)$$\bigskip

\noindent{\bf Proof.} The random variables $S^{(1)}_a$, $S^{(2)}_b$, $S^{(1)}_c$
satisfy the assumptions of Corollary (2.3) therefore
(4), holds. If also condition (5) is satisfied then, since the
variables take values in the set $\{-1,+1\}$, with probability $1$ one must 
have
$$S^{(1)}_x = - S^{(2)}_x\qquad  (x=a,b,c)\eqno(7)$$
and therefore
$\langle S^{(1)}_aS^{(1)}_c) =- \langle S^{(1)}_aS^{(2)}_c\rangle $.
Using this identity, (4) becomes (6).
\bigskip
Summing up: Theorem (1) proves that Bell's inequality is satisfied if one
takes as hypothesis the negation of his ``vital assumption''.
From this we conclude that Bell's ``vital assumption'' not only is
not ``vital'' but in fact has nothing to do with Bell's inequality.
\bigskip

\noindent{\smc Remark}. Using Lemma (14.1) below, we can allow that the
observables take values in $[-1,1]$ also in Theorem (1).\bigskip

\noindent{\smc Remark}. The above discussion is not a refutation of the Bell inequality: it is a
refutation of Bell's claim that his formulation of locality is an essential
assumption for its validity: since the locality assumption is irrelevant
for the proof of Bell's inequality it follows that this inequality
cannot discriminate between local and non local hidden variable theories, as
claimed both in the introduction and the conclusions of Bell's paper.

In particular: Theorem (1) gives an example of situations in which:

\item{(i)}  Bell's locality condition is violated while his inequality
           is satisfied.

In a recent experiment with M. Regoli [AcRe99] we have produced examples of
situations in which:

\item{(ii)}  Bell's locality condition is satisfied while his inequality
           is violated.
\bigskip

\beginsection{(6) The role of the counterfactual argument in Bell's proof }

\bigskip
Bell uses the counterfactual argument in an essential way in his proof
because it is easy to check that formula (13)
in [Bell'64] paper is the one which allows him to reduce, in the
proof of his inequality, all consideration to the $A$--variables
($S^{(1)}_a$ in our notations, while Bell's $B$--variables are
the $S^{(2)}_a$ in our notations). The pairs of chameleons (cf. section
(10), as well as the experiment of [AcRe99]
provide a counterexample precisely to this formula.

\bigskip

\beginsection{(7) Proofs of Bell's inequality based on counting arguments}

\bigskip
There is a widespread illusion to exorcize the above mentioned critiques 
by restricting one's considerations to results of measurements. 
The following considerations show why this is an illusion. 

The counting arguments, usually used to prove the Bell inequality are 
all based on the following scheme.
In the same notations used up to now, consider
$N$ simultaneous measurements of the singlet pairs of observables
$(S^1_a, S^2_b)$, $(S^2_b, S^1_c)$, $(S^2_c, S^1_a)$ and one 
denotes $S^j_{x,\nu}$ the results of the $\nu$--th measurement of
$S^j_x$ ($j=1,2$, $x=a,b,c$, $\nu=1,\dots ,N$).
With these notations one can calculate the empirical correlations on the 
samples, that is
$$ {1\over N} \sum_\nu S^1_{a,\nu}S^2_{b,\nu}=\langle S^1_aS^2_b
\rangle\eqno(1)$$
(and similarly for the other ones). 
In the Bell inequality, 3 such correlations are involved.
$$\langle S^1_aS^2_b\rangle\ ,\quad\langle S^2_bS^1_c\rangle\ ,
\quad\langle S^1_aS^2_c\rangle\eqno(2)$$
Thus in the three experiments observer $1$ has to measure
$S^1_a$ in the first and third experiment and $S^1_c$ in the second,       
while observer $2$ has to measure $S^2_b$ in the first and second experiment 
and $S^1_c$ in the third.
Therefore the directions $a$ and $b$ can be chosen arbitrarily by the
two observers and it is not necessary that observer $1$ is informed of the
choice of observer $2$ or conversely.
However the direction $c$ has to be chosen by both observers and
therefore at least on this direction there should be a preliminary
agreement among the two observers. This preliminary information 
can be replaced it by a procedure in which each observer chooses at will 
the three directions only those choices are considered for which
it happens (by chance) that the second choice of observer $1$ coincides 
with the third of observer $2$ (cf. section (15) for further discussion
of this point).
Whichever procedure has been chosen, after the results of the experiments 
one can compute the 3 empirical correlations
$$\langle S^{(1)}_aS^{(2)}_b\rangle={1\over N}\,\sum^N_{j=1}
S^{(1)}_a(p^{(1)}_j)S^{(2)}_b(p^{(1)}_j)\eqno(3)$$
$$\langle S^{(2)}_bS^{(1)}_c\rangle={1\over N}\,\sum^N_{j=1}
S^{(1)}_c(p^{(2)}_j)S^{(2)}_b(p^{(2)}_j)\eqno(4)$$
$$\langle S^{(1)}_aS^{(2)}_c\rangle={1\over N}\,\sum^N_{j=1}
S^{(1)}_a(p^{(3)}_j)S^{(2)}_c(p^{(3)}_j)\eqno(5)$$
where $p^{(3)}_j$ means the $j$--th point of the $3$--d experiment etc...
If we try to apply the Bell argument directly to the empirical data
given by the right hand sides of (3), (4), (5), we meet the expression
$${1\over N}\,\sum^N_{j=1}S^{(1)}_a(p^{(1)}_j)S^{(2)}_b(p^{(1)}_j) -
  {1\over N}\,\sum^N_{j=1}S^{(1)}_c(p^{(2)}_j)S^{(2)}_b(p^{(2)}_j)
\eqno(6)$$
from which we immediately see that, if we try to apply Bell's reasoning 
to the empirical data, we are stuck at the first step because we find a 
sum of terms of the type
$$ S^{(1)}_a(p^{(1)}_j)S^{(2)}_b(p^{(1)}_j) - 
   S^{(1)}_c(p^{(2)}_j)S^{(2)}_b(p^{(2)}_j)\eqno(7)$$
to which the inequalities among numbers, of section (1), cannot be 
applied because in general
$$ S^{(2)}_b(p^{(1)}_j) \ne S^{(2)}_b(p^{(2)}_j)\eqno(8)$$
More explicitly: since the expression (x.) above is of the form
$$ ab - b'c  $$
with $ a,b, b',c  \in \{\pm 1\}$, the only possible upper bound for it 
is $2$ and not $ 1- ac  $.

Even supposing that we, in order to uphold Bell's thesis, can
introduce a {\it cleaning operation\/} [Ac98], (cf. [AcRe99]),
which eliminates all the points in which (8) is not satisfied, we would
arrive to the inequality
$$\left| 
{1\over N}\,\sum^N_{j=1}S^{(1)}_a(p^{(1)}_j)S^{(2)}_b(p^{(1)}_j) -
  {1\over N}\,\sum^N_{j=1}S^{(1)}_c(p^{(2)}_j)S^{(2)}_b(p^{(2)}_j) \right|
\leq 1 - {1\over N}\,\sum^N_{j=1}
S^{(1)}_a(p^{(1)}_j)S^{(1)}_c(p^{(2)}_j) \eqno(9)$$
and, in order to deduce from this, something comparable with the 
experiments we need to use the counterfactual argument, assessing that
$$S^{(1)}_c(p^{(2)}_j) = -S^{(2)}_c(p^{(2)}_j) \eqno(10)$$
But in the second experiment $S^{(2)}_b $ and not $S^{(2)}_c$ has been 
measured. Thus to postulate the validity of (10) means to postulate
that: 
{\it the value assumed by $S^{(2)}_b $ in the second experiment is the same  
that we would have found if $S^{(2)}_c$ and not $S^{(2)}_b $ had been 
measured}. The chameleon effect provides a counterexample to this 
statement. \bigskip

\beginsection{(8) The quantum probabilistic analysis}

\bigskip
Given the results of section (5), (6), (7), it is then legitimate
to ask:

{\it if Bell's vital assumption is irrelevant for the deduction of
Bell's inequality, which is the really vital assumption which guarantees
the validity of this inequality?}

This natural question was first answered in [Ac81] and this result
motivated the birth of {\it quantum probability\/} as something more
than a mere noncommutative generalization of probability theory; in fact
a necessity motivated by experimental data.
\bigskip

Theorem (2.3) has only two assumptions:

\item{(i)} that the random variables take values in the interval $[-1,+1]$

\item{(ii)} that the random variables are defined on the same probability space

Since we are dealing with spin variables, assumption (i) is reasonable.

Let us consider assumption (ii). This is equivalent to the claim that
the three probability measures $P_{ab},P_{ac},P_{cb} $, representing
the distributions of the pairs $(S^{(1)}_a,S^{(2)}_b)$, $(S^{(1)}_c,S^{(2)}_b)$,
$(S^{(1)}_a,S^{(2)}_c)$ respectively, can be obtained by restriction from a single
probability measure $P$, representing the distribution of the quadruple
$S^{(1)}_a,S^{(1)}_c,S^{(2)}_b,S^{(2)}_c$.

This is indeed a strong assumption because, due to the incompatibility
of the spin variables along non parallel directions, the three
correlations
$$\langle S^{(1)}_a S^{(2)}_b\rangle  \quad , \quad  \langle  S^{(1)}_cS^{(2)}_b \rangle
\quad , \quad  \langle S^{(1)}_a S^{(2)}_c\rangle  \eqno(6)$$
can only be estimated in different, in fact mutually incompatible,
series of experiments. If we label each series of experiments by the
corresponding pair (i.e. $(a,b), (b,c), (c,a)$), then we cannot exclude
the possibility that also the probability measure in each series of
experiments will depend on the corresponding pair.
In other words, each of the measures $P_{a,b}, P_{b,c}, P_{c,a}$
describes the joint statistics of a pair of commuting observables
$(S^{(1)}_a,S^{(2)}_b)$, $(S^{(1)}_c,S^{(2)}_b)$, $(S^{(1)}_a,S^{(2)}_c)$ and there is no a priori
reason to postulate that all these joint distributions for pairs can be
deduced from a single distribution for the quadruple
$\{S^{(1)}_a,S^{(1)}_c,S^{(2)}_b,S^{(2)}_c\}$.

We have already proved in Theorem (2.3) that this strong assumption
implies the validity of the Bell inequality.
Now let us prove that it is the truly {\it vital} assumption for the validity
of this inequality, i.e. that, if this assumption is dropped, i.e. if no
single distribution for quadruples exist, then it is an easy exercise to
construct counterexamples violating Bell's inequality. To this goal one
can use the following lemma:
\bigskip
\noindent{\smc Lemma (1) \/}.
Let be given three probability measures $P_{ab},P_{ac},P_{cb} $ on a given
(measurable) space $(\Omega, {\cal F})$ and let $S_a^{(1)} , S_c^{(1)} , S_b^{(2)} , S_d^{(2)}$
be functions, defined on $(\Omega, {\cal F})$ with values in the interval
$[-1,+1]$, and such that the probability measure $P_{ab} $ (resp.
$P_{cb},P_{ac}$)  is the distribution of the pair $(S^{(1)}_a,S^{(2)}_b)$ (resp.
$(S^{(1)}_c,S^{(2)}_b)$, $(S^{(1)}_a,S^{(2)}_c)$).
For each pair define the corresponding correlation
$$\kappa_{ab}:=\langle S^{(1)}_a,S^{(2)}_b\rangle :=\int S^{(1)}_a S^{(2)}_b dP_{ab}$$
and suppose that, for $\varepsilon,\varepsilon' =\pm$, the joint
probabilities for pairs
$$P^{\varepsilon\varepsilon'}_{x,y} :=
P(S_x^{(1)}=\varepsilon  \ ;  \ S_y^{(2)}= \varepsilon') $$
satisfy:
$$P^{++}_{xy} = P^{--}_{xy} \qquad ; \qquad
P^{+-}_{xy}=P^{-+}_{xy}\eqno(1)$$
$$P^+_x=P^-_x=1/2\eqno(2)$$
then the Bell inequality
$$|\kappa_{ab}-\kappa_{bc}|\leq1-\kappa_{ac}\eqno(3)$$
is equivalent to
$$|P^{++}_{ab}-P^{++}_{bc}|+P^{++}_{ac}\leq{1\over2}\eqno(3a)$$
\noindent{\smc Proof}. The inequality (3) is equivalent to
$$|2P^{++}_{ab}
-2P^{+-}_{ab}-2P^{++}_{bc}+2P^{+-}_{bc}|\leq1-2
P^{++}_{ac}+2P^{+-}_{ac}\eqno(4)$$
Using the identity (equivalent to (2))
$$P^{+-}_{xy}={1\over2}\,-P^{++}_{xy}\eqno(5)$$
the left hand side of (4) becomes the modulus of
$$2(P^{++}_{ab}-P^{+-}_{ab})-2(P^{++}_{bc}-P^{+-}_{bc})=
2\left(P^{++}_{ab}-{1\over2}\,+P^{++}_{ab}\right)-2\left(P^{++}_{bc}
-{1\over2}\,+P^{++}_{bc}\right)$$
$$=4(P^{++}_{ab}-P^{++}_{bc})\eqno(6)$$
and, again using (5), the right hand side of (4) is equal to
$$1-2\left(P^{++}_{ac}-{1\over2}\,+P^{++}_{ac}\right)=2-4P^{++}_{ac}
\eqno(7)$$
Summing up, (3) is equivalent to
$$|P^{++}_{ab}-P^{++}_{bc}|\leq{1\over2}\,-P^{++}_{ac}\eqno(8)$$
which is (3a)
\bigskip
\noindent{\smc Corollary (2) \/}. There exist triples of
$P_{ab},P_{ac},P_{cb} $ on the $4$--point space $\{+1,-1\}\times \{+1,-1\}$
which satisfy conditions (1), (2) of Lemma (1) and are not compatible
with any probability measure $P$ on the $6$--point space
$\{+1,-1\}\times \{+1,-1\}\times \{+1,-1\}$.\bigskip

\noindent{\smc Proof}. Because of conditions (1), (2) the probability measures
$P_{ab},P_{ac},P_{cb} $ are uniquely determined by the three numbers
$$P^{++}_{ab},P^{++}_{ac},P^{++}_{cb} \in [0,1]\eqno(9)$$
Thus, if we choose these three numbers so that the inequality (3a) is
not satisfied, the Bell inequality (3) cannot be satisfied because of
Lemma (1).
\bigskip

\beginsection{(9) The realism of ballot boxes and the corresponding statistics}

\bigskip
The fact that there is no a priori reason to postulate that the joint
distributions of the pairs $(S^{(1)}_a,S^{(2)}_b)$, $(S^{(1)}_c,S^{(2)}_b)$, $(S^{(1)}_a,S^{(2)}_c)$
can be deduced from a single distribution for the quadruple
$S^{(1)}_a,S^{(1)}_c,S^{(2)}_b,S^{(2)}_c$, does not necessarily mean that such a common
joint distribution does not exist.

On the contrary, in several physically meaningful situations, we have
good reasons to expect that such a joint distribution should exist even
if it might not be accessible to direct experimental verification.

This is a simple consequence of the so--called {\it hypothesis of realism}
which is justified whenever we are entitled to believe that the results
of our measurements are {\it pre--determined}. In the words of Bell:
{\it Since we can predict in advance the result of measuring any chosen
component of $\sigma_2$, by previously measuring the same component of
$\sigma_1$, it follows that the result of any such measurement must
actually be predetermined}.

Consider for example a box containing pairs of balls. Suppose that the
experiments allow to measure either the color or the weight or the
material of which each ball is made of, but the rules of the game are
that on each ball only one measurement at a time can be performed.
Suppose moreover that the experiments show that, for each property, only
two values are realized and that, whenever a simultaneous measurement of
the same property on the two elements of a pair is performed, the
resulting answers are always discordant. Up to a change of convenction
and in appropriate units, we can always suppose that these two values are
$\pm 1$ and we shall do so in the following.

Then the joint distributions of pairs (of properties relative to
different balls) are accessible to experiment, but those of triples, or
quadruples, are not.

Nevertheless, it is reasonable to postulate that, in the box, there is a
well defined (although {\it purely Platonic}, in the sense of not being
accessible to experiment) number of balls with each given color, weight and
material. These numbers give the relative frequencies of triples of
properties for each element of the pair hence, using the perfect
anticorrelation, a family of joint probabilities for all the possible
sextuples. More precisely, due to the perfect anticorrelation, the
relative frequency of the triples of properties
$$[S^{(1)}_a= a_1] \ , \ [S^{(1)}_b =b_1]  \ , \ [S^{(1)}_c =c_ 1]$$
where $a_1 , b_1, c_ 1 =\pm 1 $ are equal to the relative frequency of the
sextuples of properties
$$[S^{(1)}_a= a_1] \ , \ [S^{(1)}_b =b_1]  \ , \ [S^{(1)}_c =c_ 1]  \ , \
[S^{(2)}_a= -a_1] \ , \ [S^{(2)}_b =-b_1]  \ , \ [S^{(2)}_c =-c_ 1]$$
and, since we are confining ourselves to the case of $3$ properties and
$2$ particles, the above ones, when $a_1 , b_1, c_ 1 $ vary in all
possible ways in the set $\{\pm 1\} $, are all the possible
configurations in this situation, {\it the counterfactural argument is
applicable\/} and in fact we have used it to deduce the joint
distribution of sextuples from the joint distributions of triples.
\bigskip

\beginsection{(10) The realism of chameleons and the corresponding statistics}

\bigskip
According to the quantum probabilistic interpretation, what Einstein,
Podolsky, Ro\-sen, Bell and several other who have discussed this topic,
call {\it the hypothesis of realism} should be called in a more precise
way the hypothesis of the {\it ballot box realism} as opposed to  hypothesis
of the {\it chameleon realism}.

The point is that, according to the quantum probabilistic interpretation, the
term {\it predetermined} should not be confused with the term {\it realized
a priori}, which has been discussed in section (9.): it might be {\it
conditionally dediced\/} according to the scheme: {\it if such and such
will happen, I will react so and so...\/}.

The chameleon provides a simple example of this distinction:
a chameleon becomes {\it deterministically} green on a leaf and brown on 
a log. In this sense we can surely claim that its color on a leaf is 
{\it predetermined}. However this does not mean that the chameleon 
was green also before jumping on the leaf.

The chameleon metaphora describes a mechanism which is {\it perfectly local},
even {\it deterministic} and surely {\it classical and macroscopic};
moreover there are no doubts that the situation it describes is {\it
absolutely realistic}. Yet this realism, being different from the {\it ballot
box realism}, allows to render free from metaphysics statements
of the orthodox interpretation such as: {\it the act of measurement
creates the value of the measured observable}. To many this looks
metaphysic or magic; but load how natural it sounds when you think of
the color of a chameleon.

Finally, and most important for its implications relatively to the EPR
argument, the chameleon realism provides a simple and natural
counterexample of a situation in which the results are {\it predetermined}
however {\bf the counterfactual argument is not applicable}.

Imagine in fact a box in which there are many pairs of chameleons. In
each pair there is exactly an healthy one, which becomes green on a
leaf and brown on a log, and a mutant one, which becomes brown on a
leaf and green on a log; moreover exactly one of the chameleons in each
pair weights $100$ grams and exactly one $200$ grams. A measurement
consists in separating the members of each pair, each one in a smaller
box, and in performing one and only one measurement on each member of
each pair.

The {\it color on the leaf, color on the log, and weight\/} are
$2$--valued observables (because we do not know a
priori if we are measuring the healthy or the mutant chameleon).
Thus, with respect to the observables: {\it color on the leaf color on
the long and weight} the pairs of
chameleons behave exactly as EPR pairs: whenever the same observable is
measured on both elements of a pair, the results are opposite. However,
suppose I measure the color on the leaf, of one element of a pair and
the weight of the other one and suppose the answers I find are: {\it green}
and {\it $100$ grams}. Can I conclude that the second element of the pair
is {\it brown and weights $100$ grams}? Clearly not because there is no
reason to believe that the second member of the pair, of which the
weight was measured while in a box, was also on a leaf.

From
this point of view the measurement interaction enters the very
definition of an observable. However also in this interpretation, which
is more similar to the quantum mechanical situation, the counterfactual
argument cannot be applied because it amounts to answer {\it ``brown''} to
the question:
{\it which is the color on the leaf, if I have measured the weight and
if I know that the chameleon is the mutant one?} (this because the
measurement of the other one gave green on the leaf). But this answer is
not correct, because it could well be that inside the box there is a
leaf and the chameleon is interacting with it while I am measuring its
weight, but it could also be that it is interacting with a log,
also contained inside the box in which case, being a mutant, it would be
green.
\bigskip
Therefore if we can produce an example of a 2-particle system in which
the Heisenberg evolution of each particle's observable satisfies Bell's
locality condition, but the Schroedinger evolution of the state, i.e.
the expectation value $\langle \cdot \rangle$, depends on the pair
$(a,b)$ of measured observables, we can claim that this counterexample
abides with the same definition of locality as Bell's theorem.
\bigskip

\beginsection{(11) Bell's inequalities and the chamaleon effect}

\bigskip
{\bf Definition (1)} Let $S$ be a physical system and ${\cal O}$ a
family of observable quantities relative to this system. We say that the
{it chamaleon effect} is realized on $S$ if, for any measurement $M$
of an observable $A\in {\cal O}$, the dynamical evolution of $S$
depends on the observable $A$. If $D$ denotes the state space of $S$,
this means that the change of state from the beginning to the end of the
experiment is described by a map (a one--parameter group or semigroup in
the case of continuous time)
$$ T_A \ : \ D \to D $$
{\bf Remark}. The explicit form of the dependence of $ T_A$ on $ A$
depends on both the system and the measurement and many concrete
examples can be constructed. An example in the quantum domain is
discussed in [Ac98] and the experiment of [AcRe99] realizes an example in
the classical domain.\bigskip

\noindent{\bf Remark} If the system $S$ is composed of two sub--systems $S_1$ and
$S_2$, we can also consider the case in which the evolutions of the two
subsystems are different in the sense that, for system $1$, we have one
form of functional dependence, $T^{(1)}_A$, of the evolution associated
to the observable $ A$ and, for system $2$, we have another form of functional
dependence, $ T^{(2)}_A$. In the experiment of [AcRe99], the state space is
the unit disk $D$ in the plane, the observables are parametrized by
angles in $[0,2\pi )$ (or equivalently by unit vectors in the unit
circle) and, for each observable $S^{(1)}_\alpha $ of system $1$
$$ T^{(1)}_\alpha := R_\alpha $$
and, for each observable $S^{(2)}_\alpha$ of system $2$
$$ T^{(2)}_\alpha := R_{\alpha+\pi} $$
where $R_{\alpha}$ denotes (counterclockwise) rotation of an angle
$\alpha$.
\bigskip
Let us consider Bell's inequalities by assuming that a chamaleon
effect
$$(S^{(1)}_a, S^{(2)}_b)\mapsto (S^{(1)}_a\circ T^{(1)}_a, S^{(2)}_b\circ T^{(2)}_b)$$
is present. Denoting $E$ the common initial state of the composite
system $(1,2)$, (e.g. singlet state), the state at the end of the
measurement will be
$$E\circ  (S^{(1)}_a\circ T^{(1)}_a, S^{(2)}_b\circ T^{(2)}_b)$$
Now replace $S^{(j)}_x$ by:
$$\tilde S^{(j)}_x:=S^{(j)}_x\circ T^{(j)}_x$$
Since the $\tilde S^{(j)}_x$ take values $\pm 1$, we know from Theorem
(2.3) that, if we postulate the existence of joint probabilities for the
triple $\tilde S^{(1)}_a , \tilde S^{(2)}_b, \tilde S^{(1)}_c$, compatible with the
two correlations $E(\tilde S^{(1)}_a \tilde S^{(2)}_b), E(\tilde S^{(1)}_c\tilde S^{(2)}_b)$,
then the inequality
$$|E(\tilde S^{(1)}_a \tilde S^{(2)}_b)-E(\tilde S^{(1)}_c\tilde S^{(2)}_b)|
\leq 1-E(\tilde S^{(1)}_a\tilde S^{(1)}_c)$$
holds and, if we also have the singlet condition
$$E(S^{(1)}_c(T^{(1)}_cp)S^{(2)}_c(T^{(2)}_cp))=-1\eqno(1)$$
then a.e.
$$\tilde S^{(1)}_c=-\tilde S^{(2)}_c$$
and we have the Bell's inequality.
Thus, if we postulate the same probability space, even the chamaleon
effect alone is not sufficient to guarantee violation of the Bell's
inequality.

Therefore the fact that the three experiments are done on different and
incompatible samples must play a crucial role.


As far as the chameleon effect is concerned, let us notice that, in the
above statement of the problem the fact that we use a single initial
probability measure $E$ is equivalent to postulate that, at time $t=0$
the three pairs of observables
$$(S^{(1)}_a , S^{(2)}_b) \quad , \quad (S^{(1)}_c, S^{(2)}_b) \quad , \quad (S^{(1)}_a, S^{(1)}_c)$$
admit a common joint distribution, in fact $E$.
\bigskip

\beginsection{(12) Physical implausibility of Bell's argument}

\bigskip
In this section we show that, combining the chameleon effect with the
fact that the three experiments refer to different samples, then even in
very simple situations, no cleaning conditions can lead to a proof of
the Bell's inequality.


If we try to apply Bell's reasoning to the empirical data, we have to
start from the expression
$$\left|{1\over N}\,\sum_j S^{(1)}_a(T^{(1)}_ap^I_j)S^{(2)}_b(T^{(2)}_bp^I_j)-{1\over
N}\,\sum_j S^{(1)}_c(T^{(1)}_cp^{II}_j)S^{(2)}_b(T^{(2)}_bp^{II}_j)\right|\eqno(1)$$
which we majorize by
$${1\over N}\,\sum_j \left|S^{(1)}_a(T^{(1)}_ap^I_j)S^{(2)}_b(T^{(2)}_bp^I_j)-
S^{(1)}_c(T^{(1)}_cp^{II}_j)S^{(2)}_b(T^{(2)}_bp^{II}_j)\right|\eqno(2)$$
But, if we try to apply the inequality among numbers to the expression
$$\left|S^{(1)}_a(T^{(1)}_ap^I_j)S^{(2)}_b(T^{(2)}_bp^I_j)-
S^{(1)}_c(T^{(1)}_cp^{II}_j)S^{(2)}_b(T^{(2)}_bp^{II}_j)\right|\eqno(3)$$
we see that we are not dealing with the situation covered by Corollary
(1.2), i.e.
$$|ab - cb|\leq 1 - ac\eqno(4)$$
because, since
$$S^{(2)}_b(T^{(2)}_bp^I_j) \ne S^{(2)}_b(T^{(2)}_bp^{II}_j) \eqno(5)$$
the left hand side of (4) must be replaced by
$$|ab - cb'|\eqno(6)$$
whose maximum, for $a,b, c,b'\in [-1,+1]$ is $2$ and not $1 - ac$.

Bell's implicit assumption of the single probability space is equivalent to
the postulate that, for each $j=1, \dots ,N$
$$p^I_j=p^{II}_j\eqno(7)$$
Physically this means that:

{\it the hidden parameter in the first experiment is the same as the
hidden parameter in the second experiment}

This is surely a very implausible assumption.

Notice however that, without this assumption, Bell's argument cannot be
carried over and we cannot deduce the inequality because we must stop at
equation (2).
\bigskip

\beginsection{(13) The role of the single probability space in CHSH's proof}

\bigskip
Clauser, Horne, Shimony, Holt [ClHo69] introduced the variant (2.6)
of the Bell inequality for quadruples
$(a,b)$, $(a,b')$, $(a',b)$, $(a',b')$ which is based on
the following inequality among numbers
$$\mid ab+ab'+a'b-a'b'\mid \leq 2 \eqno(1)$$
Section (1) already contains a proof of (1).
For $ a,b,b',a \in [-1,1]$, a direct proof follows from
$$\mid b+b'\mid + \mid b-b'\mid \leq 2\eqno(2)$$
because
$$\mid ab+ab'+a'b-a'b'\mid =\mid a(b+b')+a'(b-b')\mid  \leq
\mid a\mid \cdot \mid b+b'\mid +\mid a'\mid \cdot \mid b-b'\mid
\leq \mid b+b'\mid +\mid b-b'\mid  \leq 2$$
The proof of (2) is obvious because it is equivalent to
$$\mid b+b'\mid^2 + \mid b-b'\mid^2 =b^2+b'^2 + 2bb' + b^2 + b'^2 -2bb'
=2b^2+2b'^2 \leq 4$$
which is identically satisfied (cf. also Lemma (1.1)).
\bigskip
{\bf Remark (1)} Notice that an inequality of the form
$$\mid a_1b_1+a_2b_2'+a_3'b_3-a_4'b_4'\mid \leq 2 \eqno(3)$$
would be obviously false. In fact, for example the choice
$$a_1=b_1=a_2=b_2'=a_3'=b_3=b'_4 =1 \qquad ; \qquad a'_4=-1$$
would give
$$\mid a_1b_1+a_2b_2'+a_3'b_3-a_4'b_4'\mid = 4$$
That is: for the validity of (1) it is absolutely essential that the
number $a$ is the same in the first and the second term and similarly
for $a'$ in the 3--d and the 4--th, $b'$ in the 2--d and the 4--th,
$b$ in the first and the 3--d.
\bigskip

This inequality among numbers can be extended to pairs of random
variables by introducing the following postulates:

\item{\bf (P1)} Instead of four numbers $ a,b,b',a \in [-1,1] $, one considers
four functions
$$S_a^{(1)},S_b^{(2)}, S_{a'}^{(1)},S_{b'}^{(2)}$$
all defined on the same space $\Lambda$ (whose points are called {\it hidden parameters})
and with values in $[-1,1] $.
\item{\bf (P2)} One postulates that there exists a probability measure
$P$ on $\Lambda$ which defines the joint distribution of each of the
following four pairs of functions
$$(S_a^{(1)},S_b^{(2)}), (S_a^{(1)},S_{b'}^{(2)}), 
(S_{a'}^{(1)},S_b^{(2)}),
(S_{a'}^{(1)},S_{b'}^{(2)}) \eqno(4)$$
{\bf Remark (2)} Notice that $(P2)$ automatically implies that the joint
distributions of the four pairs of functions can be deduced from a joint
distribution of the whole quadruple, i.e. the existence of a single
Kolmogorov model for these four pairs.
\bigskip
With these premises, for each $\lambda\in \Lambda$ one can apply the
inequality (1) to the four numbers
$$S_a^{(1)}(\lambda),S_b^{(2)}(\lambda), S_{a'}^{(1)}(\lambda),
S_{b'}^{(2)}(\lambda)$$
and deduce that
$$\mid S_a^{(1)}(\lambda)S_b^{(2)}(\lambda) +
       S_a^{(1)}(\lambda)S_{b'}^{(2)}(\lambda) +
       S_{a'}^{(1)}(\lambda)S_b^{(2)}(\lambda)
     - S_{a'}^{(1)}(\lambda)S_{b'}^{(2)}(\lambda) \mid \leq 2\eqno(5)$$
From this, taking $P$--averages, one obtains
$$\mid \langle S_a^{(1)}S_b^{(2)}\rangle  +
       \langle S_a^{(1)}S_{b'}^{(2)}\rangle  +
       \langle S_{a'}^{(1)}S_b^{(2)}\rangle
     - \langle S_{a'}^{(1)}S_{b'}^{(2)}\rangle  \mid =\eqno(6a)$$
$$\mid \int \Bigl( S_a^{(1)}(\lambda)S_b^{(2)}(\lambda) +
            S_a^{(1)}(\lambda)S_{b'}^{(2)}(\lambda) +
            S_{a'}^{(1)}(\lambda)S_b^{(2)}(\lambda)
     -      S_{a'}^{(1)}(\lambda)S_{b'}^{(2)}(\lambda) \Bigr)dP(\lambda)
\mid \leq \eqno(6b)$$
$$\leq \int \mid  S_a^{(1)}(\lambda)S_b^{(2)}(\lambda) +
            S_a^{(1)}(\lambda)S_{b'}^{(2)}(\lambda) +
            S_{a'}^{(1)}(\lambda)S_b^{(2)}(\lambda)
     -      S_{a'}^{(1)}(\lambda)S_{b'}^{(2)}(\lambda) \mid dP(\lambda)
\leq 2 \eqno(6c)$$\bigskip

\noindent{\bf Remark (3)} Notice that in the step from (6a) to (6b) we have used
in an essential
way the existence of a joint distribution for the whole quadruple, i.e.
the fact that all these random variales can be realized in the
same probability space.
\bigskip
In EPR type experiments we are interested in the case in which the four
pairs $(a,b)$, $(a,b')$, $(a',b)$, $(a',b')$ come from four mutually incompatible
experiments. Let us assume that there is a hidden parameter, determining
the result of each of these experiments. This means that we interpret
the number $S_a^{(1)}(\lambda)$ as the value of the spin of particle $1$
in direction $a$, determined by the hidden parameter $\lambda$.

There is obviously no reason to postulate that the hidden
parameter, determining the result of the first experiment is exactly the
same one which determines the result of the second experiment.
However, when CHSH consider the quantity (5), they are implicitly doing
the much stronger assumption that the same hidden parameter $\lambda$
determines the results of all the four experiments. This assumption is
quite unreasonable from the physical point of view and in any case it
is a much stronger assumption than simply postulating the existence of hidden
parameters. The latter assumption would allow CHSH only to consider the
expression
$$S_a^{(1)}(\lambda_1)S_b^{(2)}(\lambda_1) +
       S_a^{(1)}(\lambda_2)S_{b'}^{(2)}(\lambda_2) +
       S_{a'}^{(1)}(\lambda_3)S_b^{(2)}(\lambda_3)
     - S_{a'}^{(1)}(\lambda_4)S_{b'}^{(2)}(\lambda_4) \eqno(4)$$
and, as shown in Remark (1.) above the maximum of this expression is not $2$
but $4$ and this does not allow to deduce the Bell inequality.
\bigskip

\beginsection{(14) The role of the counterfactual argument in CHSH's proof }

\bigskip
Contrarily to the original Bell's argument, the CHSH proof of the Bell
inequality does not use explicitly the counterfactual argument.
Since one can perform experiments also on quadruples, rather than on
triples, as originally proposed by Bell, has led some authors to claim
that the counterfactual argument is not essential in the deduction
of the Bell inequality.
However we have just seen in section (7.) that the hidden assumption as
in Bell's proof, i.e. the realizability of all the random variales
involved in the same probability space, is also present in the CHSH
argument. The following lemma shows that, under the singlet assumption,
the conclusion of the counterfactual argument follows from the hidden
assumption of Bell and of CHSH.
\bigskip
{\bf Lemma (1)}
If $f$ and $g$ are random variables defined on a probability space 
$(\Lambda , P)$ and with values in $[-1,1]$, then
$$ \langle fg\rangle  :=  \int_\Lambda fg dP =-1$$
if and only if  
$$P(fg  = -1)=1$$
{\bf Proof. }  If $P(fg>-1)  >0$, then
$$ \int_\Lambda fg dP = - P(fg=-1)  - \int_{fg >-1 } |fg| dP >
- P(fg=-1)  - P(fg>-1)  >-1$$
\bigskip
{\bf Corollary (2)} Suppose that all the random variales in (x.3) are realized
in the same probability space. Then, if the singlet condition:
$$ \langle S_{x}^{(1)} S_{x}^{(2)}\rangle  = - 1 \eqno(1)$$
is satisfied, then the condition
$$ S_{x}^{(1)} = - S_{x}^{(2)} \eqno(2)$$
(i.e. formula (13) in Bell's '64 paper) is true almost
everywhere.\bigskip

\noindent{\bf Proof. } Follows from Lemma (1) with the choice
$f= S_{x}^{(1)} $, $ g=S_{x}^{(2)}$.
\bigskip
Summing up: if you want to compare the predictions of a hidden variable
theory with quantum theory in the EPR experiment (so that at least we admit
the validity of the singlet law) then the hidden assumption, of
realizability of all the random variables in (3) in the same probability space,
(without which Bell's inequality cannot be proved) implies the same
conclusion of the counterfactual argument. Stated otherwise: the
counterfactual argument is implicit when you postulate the singlet
condition and the realizability on a single probability space. It does
not matter if you use triples or quadruples.
\bigskip

\beginsection{(15) Physical difference between the CHSH's and the original
Bell's inequalities}

\bigskip
In the CHSH scheme:
$$(a,b)\ ,\quad(a',b')\ ,\quad(a,b')\ ,\quad(a',b')$$
the agreement required by the experimenters is the following:

-- $1$ will measures the same observable in experiments I and III, and the
same observable in experiments II and IV;

-- $2$ will measure the same observable in experiments I and II, and the
same observable in experiments III and IV.

Here there is no restriction a priori on the choice of the observables
to be measured.

In the Bell scheme the experimentalists agree that:

-- $1$ measures the same observable in experiments I and III,

-- $2$ measures the same observable in experiments I and II

-- $1$ and $2$ choose a priori, i.e. before the experiment begins, a
direction $c$ and agree that $1$ will measure spin in direction $c$ in
experiment II and  $2$ will measure spin in direction $c$ in
experiment III (strong agreement)

The strong agreement can be replaced by the following (weak agreement):

-- $1$ and $2$ choose a priori, i.e. before the experiment begins, a
finite set of directions $c_1, \dots , c_K$ and agree that $1$ will measure
spin in a direction choosen randomly among the directions $c_1, \dots , c_K$
in experiment II and  $2$ will do the same in experiment III

In this scheme there is an a priori restriction on the choice of some of the
observables to be measured.

If the directions, fixed a priori in the plane, are $K$, then the probability
of a coincidence, corresponding to a totally random (equiprobable)
choice, is
$$P(x^{(1)}_{II}=x^{(2)}_{III})=\sum^K_{\alpha=1}(x^{(1)}_{II}=
\alpha;x^{(2)}_{III}=\alpha)=\sum^K_{\alpha=1}{1\over K^2}\,={1\over K}$$

This shows that, contrarily than in the CHSH scheme, the choice has to be
restricted to a finite number of possibilities otherwise the probability
of coincidence will be zero.

From this point of view we can claim that the Clauser, Horne, Shimony,
Holt formulation of Bell's inequalities realize an improvement with
respect to the original Bell's formulation.\bigskip

\beginsection{Bibliography}

\bigskip
  \item{[Ac81]} 
  Luigi Accardi:
``Topics in quantum probability'', Phys. Rep. 77 (1981) 169-192

\item{[Ac97]}  Luigi Accardi:
Urne e camaleonti.
Dialogo sulla realt\`a, le leggi del caso e la teoria quantistica.
Il Saggiatore (1997).
Japanese translation, Maruzen (2000), russian translation, ed. by Igor 
Volovich, PHASIS Publishing House (2000), english translation by Daniele 
Tartaglia, to appear  

\item{[Ac99]} 
Luigi Accardi:  
On the EPR paradox and the Bell inequality
Volterra Preprint (1998) N. 350.

\item{[AcRe99a] }
Luigi Accardi, Massimo Regoli:
Quantum probability and the interpretation of quantum mechanics: a crucial
experiment, 
Invited talk at the workshop: {\it``The applications of mathematics
to the sciences of nature: critical moments and aspetcs''\/}, Arcidosso
June 28-July 1 (1999). To appear in the proceedings of the workshop, Preprint 
Volterra N. 399 (1999)

\item{[AcRe99b]}
Luigi Accardi, Massimo Regoli:
Local realistic violation of Bell's inequality: an experiment,
Conference given by the first--named author at the Dipartimento di Fisica, 
Universit\`a di Pavia on 24-02-2000, Preprint Volterra N. 402

\item{[AcRe00]}
Luigi Accardi, Massimo Regoli:
Non--locality and quantum theory: new experimental evidence,
Invited talk given by the first--named author at the Conference:  
``Quantum paradoxes'', University of Nottingham, on 4-05-2000,
Preprint Volterra N. 421

\item{[Be64]}
Bell J.S:
On the Einstein Podolsky Rosen Paradox
Physics 1 no.3. 195-200 1964.

\item{[Be66]}
Bell J.S.:
On the Problem of Hidden Variables in Quantum Mechanics.
Rev. Mod. Phys. 38 (1966) 447-452

\item{[ClHo69]}
J.F. Clauser , M.A. Horne, A. Shimony, R. A. Holt,
Phys. Rev. Letters, {\bf 49}, 1804-1806 (1969);
J. S. Bell, {\it Speakable and unspeakable in quantum mechanics.}
(Cambridge Univ. Press, 1987).

\item{[ClHo74]}
Clauser J.F., Horne M.A.:
Experimental Consequences of Objective Local Theories.
Physical Review D, vol. 10, no. 2 (1974)

\item{[EPR35]}
Einstein A., Podolsky B., Rosen N.
Can quantum mechanical description of
reality be considered complete ?
Phys. Rev. 47 (1935) 777-780

\item{[Ein49]}
A. Einstein in: Albert Einstein: Philosopher Scientist.
Edited by P.A. Schilpp, Library of Living Philosophers, Evanston,
Illinois, p.85 (1949)\end